\documentclass[useAMS]{mn2e}

\def\Msol{M_\odot}
\def\mnras{MNRAS}
\def\apj{ApJ
}\def\apjl{ApJL}
\def\apjs{ApJS}
\def\aap{A\& A}
\def\pasp{PASP}
\def\pasj{PASJ}
\def\nat{Nature}
\usepackage{times}
\usepackage{amssymb}
\usepackage{rotating}
\input{epsf}

\title[Measuring spin in AGN]
{A new way to measure supermassive black hole spin in accretion disc dominated Active Galaxies}

\author[C. Done, C. Jin, M. Middleton, M. Ward ]
{Chris Done$\thanks{E-mail:chris.done@durham.ac.uk}^1$, 
  C. Jin$^{1,2}$,  M. Middleton$^{1,3}$, Martin Ward$^1$
\\
$^1$ Department of Physics, University of Durham, South Road,
Durham DH1 3LE, UK\\
$^2$ Key Laboratory for Particle Astrophysics, Institute of High Energy Physics, CAS, 19B Yuquan Road, Beijing 100049, China.\\
$^3$ Astronomical Institute Anton Pannekoek, Science Park 904, 1098 XH Amsterdam, The Netherlands\\
}

\date{Submitted to MNRAS}
\pagerange{\pageref{firstpage}--\pageref{lastpage}} \pubyear{2009}

\begin{document}

\topmargin = -0.5cm

\maketitle

\label{firstpage}

\begin{abstract}

We show that disc continuum fitting can be used to constrain black
hole spin in a subclass of Narrow Line Seyfert 1 (NLS1) AGN as their
low mass and high mass accretion rate means that the disc peaks at
energies just below the soft X-ray bandpass. We apply the technique to
the NLS1 PG1244+026, where the optical/UV/X-ray spectrum is consistent
with being dominated by a standard disc component. This gives a best
estimate for black hole spin which is low, with a firm upper limit of
$a_*<0.86$.  This contrasts with the recent X-ray determinations of
(close to) maximal black hole spin in other NLS1 based on relativistic
smearing of the iron profile.  While our data on PG1244+026 does not
have sufficient statistics at high energy to give a good measure of
black hole spin from the iron line profile, cosmological simulations
predict that black holes with similar masses have similar growth
histories and so should have similar spins.  This suggests that there
is a problem either in our understanding of disc spectra, or/and X-ray
reflection or/and the evolution of black hole spin.

\end{abstract}


\section{Introduction} \label{sec:introduction}

'Black holes have no hair', meaning that they have no distinguishing
features, other than mass and spin (charge is negligible in an
astrophysical setting). However, they are most easily observed via a
luminous accretion flow, so mass accretion rate is another important
quantity in determining their appearance, with a weak dependence on
inclination angle. Thus there are only four parameters, and yet there
is a wide diversity in the observed properties of AGN. While the black
hole mass and mass accretion rate can be reasonably well determined, spin
only leaves an imprint on the space-time close to the event horizon,
so is hard to measure. Hence it is often the first candidate to
explain any property which is not well understood, such as the
emergence of powerful radio jets (e.g. Begelman, Blandford \& Rees
1984), feedback from which controls the star formation powered growth
of galaxies (e.g. Bower et al 2006). However, spin has wider
importance as it preserves the history of how the mass of the black
hole grows over cosmic time (Volonteri et al 2005; Fanidakis et al
2011), determines the gravitational wave signature arising from black
hole coalescence in galaxy mergers (Centrella et al 2010), and whether
the resulting black hole is likely to be ejected from the host galaxy
(King et al 2008).

Currently, the only well established method to measure black hole spin
in Active Galactic Nuclei (AGN) is from the iron K$\alpha$ line
profile.  X-ray illumination of the accretion disc gives rise to a
fluorescent iron K$\alpha$ line at 6.4-7 keV and associated continuum
reflection, both of which are sculpted by special and general
relativistic effects. Larger line widths require material closer to
the black hole, and hence imply higher spin (Fabian et al 1989; 2000).

A range of black hole spins are found with this technique (e.g the
compilations of Nandra et al 2007; Brenneman \& Reynolds 2009; de la
Calle Perez et al 2010), but a subset of low luminosity spectra from
the most variable Narrow Line Seyfert 1 (NLS1) galaxies (Gallo 2006)
show dramatically broad iron features. These require extreme spin
if these are produced primarily through relativistic reflection
e.g. MCG-6-30-15 (Wilms et al 2001; Fabian et al 2002; Brenneman \&
Reynolds 2006; Miniutti et al 2007; Chiang \& Fabian 2011), and
1H0707-495 (Fabian et al 2004; 2009).  However, the models developed
to explain these low luminosity datasets are extreme, not just in terms
of black hole spin, but also in requiring a source geometry where the
continuum can be strongly gravitationally focussed onto the inner disc
to give reflection dominated spectra and an extremely centrally
concentrated emissivity (Fabian et al 2004; Miniutti \& Fabian 2004;
Fabian et al 2009; Zoghbi et al 2010).

These extreme parameters motivated alternative models where the
spectral curvature around the iron line is instead produced by
complex absorption (e.g. Mkn 766: Turner et al 2007; Miller et al
2007; MCG-6-30-15: Miller et al 2009). The broad features are then
indicative of winds and feedback from the AGN rather than black hole
spin (Sim et al 2010; Tatum et al 2012). 

There is hope that the controversy over the nature of the spectra may
be settled using new combined spectral-timing analysis techniques
(Fabian et al 2009; Wilkins et al 2013 but see Legg et al 2012) and/or
new high energy data from NuStar (Risaliti et al 2013 but see Miller
\& Turner 2013). However, the current debate even on these topics
highlights the need for another method to
measure of black hole spin. One possibilitity is from the soft X-ray
excess, an additional component seen ubiquitously in AGN alongside the
expected disc and high energy power-law tail. If this is also formed
from reflection, then black hole spin can be constrained by the amount
of relativistic smearing of the soft X-ray lines as well as the iron
K$\alpha$ line (Crummy et al 2006). However, this remains
controversial, as the soft X-ray excess may instead be a separate
continuum component (see Section 2, below).

Instead, there is another well studied method to measure black hole
spin which is routinely applied to Galactic black holes binaries
(BHB). These can have accretion disc spectra which peak at X-ray energies,
as expected from standard disc models (Shakura \& Sunyaev 1973). A
single measure of the peak temperature, $T_{max}$, and total disc
luminosity, $L$, give an estimate of black hole spin when the system
parameters (black hole mass, inclination and distance) are known. This
derived size scale is observed to remain constant despite large
changes in mass accretion rate, giving confidence in the standard disc
models (Ebisawa et al 1991; 1993; Kubota et al 2001, Gierlinski \&
Done 2004; Davis, Done \& Blaes 2006; Steiner et al 2010). However,
this constant radius (i.e. $L\propto T_{max}^4$ behaviour is only seen
when the disc dominates the spectrum. The reconstruction of the disc
intrinsic luminosity and temperature becomes progressively more model
dependent where the high energy tail contributes more than $\sim 20$\%
of the bolometric luminosity, hence such data are not reliable
estimators of black hole spin (Kubota et al 2001; Kubota \& Done 2004;
Steiner et al 2010).

This technique has not been widely used in AGN, predominantly because
the predicted disc spectra depend on both black hole mass and mass
accretion rate, so without a good mass estimate it is not possible to
accurately determine the position of the peak disc emission. However,
generically this peak should lie in the UV region, which cannot be
directly observed in low redshift AGN due to interstellar absorption,
so there was no strong motivation to study this further. AGN spectra are
also generally not dominated by the thermal disc component, but have
substantial luminosity at higher energies (Elvis et al 1994; Richards
et al 2006) i.e. where BHB show that spin determination from the disc
continuum component is not robust.

There are three key factors which allow us to now apply this technique
to some AGN to constrain their black hole spin.  Firstly, black hole
mass in AGN can now be estimated via scaling relationships
based on the optical broad line region widths (e.g Kaspi et al
2000). The optical continuum from the disc then directly measures the
mass accretion rate through the outer disc (with a weak dependence on
inclination: Davis \& Laor 2011). Secondly, we have recently
identified a new class of AGN, whose spectra are dominated by the disc
(Jin et al 2012a; b hereafter J12a,b and Done et al 2012; hereafter
D12; Terashima et al 2012). These do have a high energy tail and a
soft X-ray excess, but the luminosity in these components is small
compared to the disc emission. Hence they form a subset of objects
where disc continuum fitting model can be used with some
confidence. These objects are all NLS1, so have low mass black holes,
and high mass accretion rates (Boroson 2002). This combination gives
the highest predicted disc temperatures, peaking in the EUV rather
than the UV, so increasing spin leads to the Wien tail of the disc
emission extending into the observable soft X-ray bandpass
(D12). Thirdly, we have developed new improved disc models which
approximately incorporate the results of full radiative transfer
through the disc photosphere via a colour temperature correction to
the blackbody temperature (D12). Such models have previously only been
widely available for spectral fitting stellar mass black holes in
binary systems (Li et al 2005).

We demonstrate the technique using PG 1244+026, a bright, low redshift
(z = 0.048, corresponding to D = 211 Mpc) disc-dominated NLS1 AGN
(J12a,b, see also Figs
\ref{fig:pg1244_opt} and \ref{fig:pg1244_conv}), where absorption
corrections to both the UV and soft X-ray emission are small due to
the low column along the line of sight.  This is not one of the NLS1
which shows a low flux state, so does not have the extreme iron line
features, but instead has a relatively simple X-ray continuum shape
(see Figs. \ref{fig:pg1244_opt} and \ref{fig:pg1244_conv}).  We use a new,
100ks, high quality dataset from the XMM-Newton satellite (see Jin et
al 2013, hereafter J13) to improve
statistics over those of J12a,b, and to determine the black hole mass
via X-ray variability as well as the H$\beta$ line width. Using the
maximum mass produces a lower limit on the disc temperature which
strongly requires low spin in order not to overpredict the observed
soft X-ray flux. Extending the AGN continuum model of D12 to include
relativistic effects, inclination dependence and advection does not
substantially change this conclusion.

A low spin for this NLS1 is in sharp contrast with the high spin derived
for other NLS1 which show low flux episodes described above. Our X-ray
data are not sufficient to derive the profile of the iron line with
high confidence, so we cannot yet say whether the low spin
determination in this object is in conflict with its iron line
profile.  Better high energy data are required in order to determine
whether this new method gives a spin estimate which is consistent with
that derived from the iron line, or whether it instead reveals a lack
of understanding of disc continuum emission and/or of disc reflection. 

\begin{figure*}
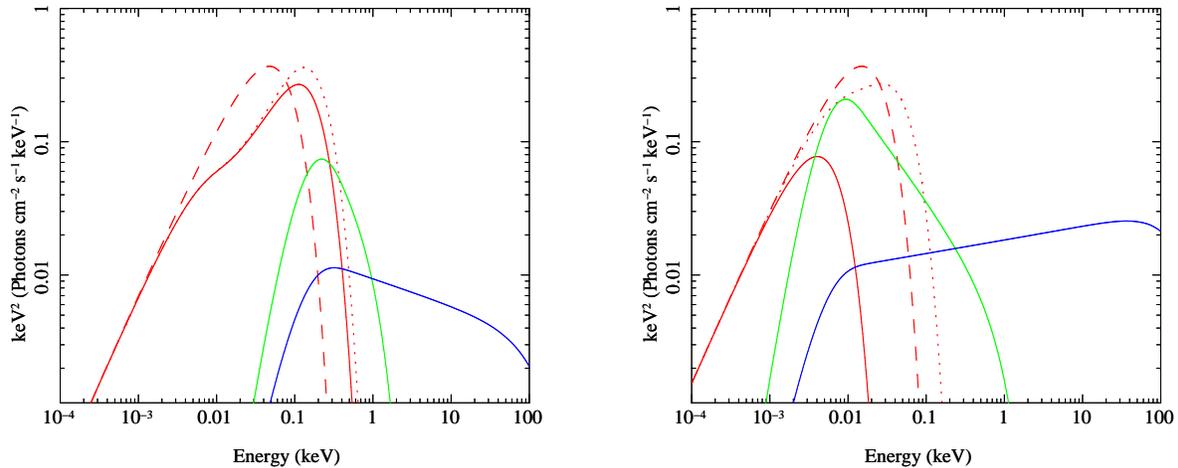

\centering
\begin{tabular}{cc}
\leavevmode  
\epsfxsize=0.45\textwidth \epsfbox{nls1_tot.ps} &
\epsfxsize=0.45\textwidth \epsfbox{bls1_tot.ps}\\
\end{tabular}
\caption{Typical accretion flow spectra for Schwarzschild black holes
with a (colour temperature corrected) standard disc (solid red line)
extending down to a radius $R_{cor}$, below which 70\% of the
accretion flow energy is emitted as a soft X-ray excess, modelled as
an additional cool, optically thick Comptonisation component with
$kT_e=0.2$~keV and $\tau=15$ (green solid line) while the remaining
30\% powers a hot Comptonisation component with $kT_e=100$~keV. The
dotted red line shows the (colour temperature corrected) standard disc
emission for $R_{cor}=6R_g$, while the dashed red line shows the
effect of removing the colour temperature correction. a) NLS1
$10^7\Msol$, $L/L_{Edd}=1$ with $R_{cor}=15R_g$ and
$\Gamma=2.2$ and b) $10^8\Msol$, $L/L_{Edd}=0.1$ with
$R_{cor}=100R_g$ and $\Gamma=1.9$. 
}
\label{fig:optx_range}
\end{figure*}

\section{Disc models and the nature of the soft X-ray excess}

Since the early 1980s it was recognised that the blue optical/UV
continuum from AGN is from a geometrically thin, optically thick
accretion disc around a supermassive black hole (Malkan 1983). This
disc emission should not completely thermalise, as electron scattering
in the disc is important as well as true absorption (e.g. Czerny \&
Elvis 1987; Ross, Fabian \& Mineshige 1991). The photospheric emission
from the disc including these effects can be calculated from full
radiative transfer models (e.g. the TLUSTY code of Hubeny et al 2001,
as used by Davis \& Laor 2011). However, in the stellar mass BHB,
these effects are typically modelled as a colour temperature
correction, $f_{col}$ where the local flux $f_\nu(T)=f_{col}^{-4}
B_\nu(f_{col}T)$, where $T$ is the predicted blackbody temperature at
radius $r$.

For the $10\Msol$ BHB, typical values are $f_{col}\sim 1.8$ (Shimura
\& Takahara 1995). Observations of LMC X-3 at $L/L_{Edd}\sim 0.5$ show
a disc dominated spectrum which peaks in $\nu f_\nu$ at $\sim 3.5$~keV
(Kolehmainen \& Done 2013). Scaling this by a factor of
$(10^6)^{-1/4}$ as appropriate for a $10^7\Msol$ at $L/L_{Edd}\sim
0.5$ predicts that low mass, high mass accretion rate AGN such as NLS1
should have discs which peak at $0.1$~keV, close to the observable
soft X-ray range. The potential detectablility is enhanced by the
slightly larger colour temperature expected in AGN ($f_{col}\sim
2-2.5$, D12) together with the higher mass accretion rates
($L/L_{Edd}\sim 1$) of most NLS1.

While the disc emission is relatively well understood, it is accompanied in
both BHB and AGN by coronal emission to much higher energies whose
origin is much less clear. AGN spectra also typically show an
additional component, termed the soft X-ray excess. This appears as a
smoothly rising continuum component over the extrapolated 2-10~keV
power law coronal emission below 1~keV. It can be well fit by cool,
optically thick, thermal Comptonisation emission in addition to the
hot, optically thin Compton emission required to make the 2-10~keV
power law. However, the temperature of this cool component remains
remarkably constant despite changes in the predicted underlying
accretion disc temperature, making this solution appear fine tuned
(Czerny et al 2003; Gierlinski \& Done 2004).

Another model for the origin of the soft X-ray excess is that it is
produced by reflection from partially ionised material. The reduced
absorption opacity below the ionised Oxygen edge at $\sim 0.7$~keV
gives an increased reflectivity at low energies, producing excess
emission over the intrinsic power law at these energies (Ross \&
Fabian 1993; Ross \& Fabian 2005).  This has the advantage that it
produces the soft excess at a fixed energy as observed, though only
for a fixed ionisation parameter which equally requires fine tuning
(Done \& Nayakshin 2007). This ionisation state includes strong line
emission from iron L and Oxygen which are not seen as narrow features
in the data, so extreme relativistic effects (high black hole spin 
and centrally focussed emissivity)
are required to smear this reflected emission into a smooth continuum
(Crummy et al 2006). The shape of the soft X-ray excess can then be
used as an independent tracer of black hole spin.

The similarity of extreme relativistic effects required to explain
both the iron line and soft X-ray excess continuum is used as an
argument that both are indeed formed in this way (Crummy et al
2006). However, this is also currently controversial, with
counterexamples where different spins are required to explain the soft excess
and iron line profiles (Patrick et al 2011), though this mismatch may
also point to multiple reflectors with similar relativisitic effects
but different ionisation parameters (e.g. Fabian et al 2004).

However, there is now growing evidence from variability studies that
the soft excess represents a true additional component connected to
the disc rather than to the high energy power law. A monitoring campaign on
Mkn509 (a standard broad line Seyfert 1, hereafter BLS1) shows that
the long term variability of the soft excess correlates with that of
the UV but not with the hard X-rays (Mehdipour et al 2011).
Conversely, on short timescales, the soft excess remains constant
while the hard X-rays vary (Noda et al 2011; 2013). Similar lack of
soft X-ray variability with strong hard X-ray variability is also seen
in the disc dominated NLS1 (RE J1034+396: Middleton et al 2009; RX
J0136.9-3510: Jin et al 2009), including this object (PG1244+026: Jin
et al 2013, hereafter J13). This all suggests that the
soft excess in these objects is again dominated by a true additional
component which links to the UV disc rather than to the power law. We
note that an alternative, reflection dominated interpretation of the
spectrum of RX J1034+396 has both the 0.5-0.7 and 5-10~keV spectra
dominated by a single reflection component, so cannot explain the very
different amounts of variability seen at these energies (Zoghbi \&
Fabian 2011; their Fig 6, see discussion in J13).

\begin{table*}
\begin{tabular}{lc|l|l|l|l|l|l|l|l}
 \hline
   \small $N_H$ ($10^{20}$~cm$^{-2}$) & 
$\log L/L_{Edd}$ & $r_{corona}$ ($R_g$)&
$kT_e$ (keV)& $\tau$ & $\Gamma$ & $f_{pl}$ & $\chi^2/\nu$ \\ 
\hline
$2.9\pm 0.2$ & $-0.29\pm 0.01$ & $17.3\pm 0.3$ &
$0.22\pm 0.01$ & $15\pm 0.5$ & $2.28\pm 0.02$ & $0.24\pm 0.01$ & 1972/1068\\
\hline
\end{tabular}
\caption{Details of the {\sc optxagnf} fit to PG1244+026 shown in 
Fig \ref{fig:pg1244_opt}. The
remaining paramaters of black hole mass and spin 
are fixed at $2\times 10^7 M_\odot$ and  $a_*=0$, respectively.
}
\label{tab:pg_optxagnf}
\end{table*}

Nonetheless, the low state spectra of the extreme variability
subclass of NLS1 may contain substantially more reflected emission
(e.g.  Fabian et al 2009). These objects do indeed show a
significantly different
pattern of X-ray variability (Gierlinski \& Done 2006), although there
is probably a contribution from an additional component at the softest
energies even in these objects (Zoghbi, Uttley \& Fabian 2011).

Whatever the origin of the soft X-ray excess, and the higher energy
coronal emission, together they can carry a substantial fraction of
the accretion power (e.g. the standard quasar template spectra of
Elvis et al 1994; Richards et al 2006 and PG1048+213 in D12). If this
is powered by the same accretion flow as powers the UV disc emission
then energy conservation requires that part of the accretion energy
must be dissipated instead in a non-standard disc component. We have
developed a model, {\sc optxagnf}, which incorporates conservation of
energy by assuming that the accretion flow thermalisises to a (colour
temperature corrected) blackbody at radii larger than $R_{cor}$ but
that below this the disc density becomes too low for thermalisation
(perhaps because its scale height increases due to a UV
line/radiatively driven disc wind or perhaps due to this wind failing
to escape and impacting back on the surface of the disc and giving
rise to shock heating of the photosphere: e.g streamlines in Risaliti
\& Elvis 2010). Some fraction, $f_{pl}$, of the gravitational energy
within $R_{cor}$ is emitted as the standard high energy coronal
component while the remainder $(1-f_{pl})$ of the energy forms the
soft X-ray excess, modelled as a cool, optically thick, thermal
Comptonisation component (D12).

Figure \ref{fig:optx_range}a shows typical model spectra for a NLS1 ($10^7\Msol$,
$L/L_{Edd}=1$). The red dotted line shows a standard accretion disc
spectrum assuming complete thermalisation down to the last stable
orbit of a non-spinning black hole, while the dashed line includes the
colour temperature correction from electron scattering in the disc
photosphere as a function of radius. The red solid line shows the disc
spectrum truncated at $R_{cor}=15R_g$, with 30\% ($f_{pl}=0.3$) of the
remaining energy dissipated in a hot ($kT_e=100$~keV), optically thin
corona with photon spectral index $\Gamma=2.2$ (blue), with the rest forming 
the soft X-ray excess from a cool ($kT_e=0.2$~keV), optically thick
($\tau_T=15$) region (green).

Figure \ref{fig:optx_range}b shows the same sequence of spectra for a
BLS1 ($10^8\Msol$, $L/L_{Edd}=0.1$). All other parameters are kept
constant except that $R_{cor}=100R_g$ and $\Gamma=1.9$. Figs
\ref{fig:optx_range}a,b span the range of observed spectral energy
distributions seen in J12a,b.

\begin{figure}
\epsfxsize=0.5\textwidth \epsfbox{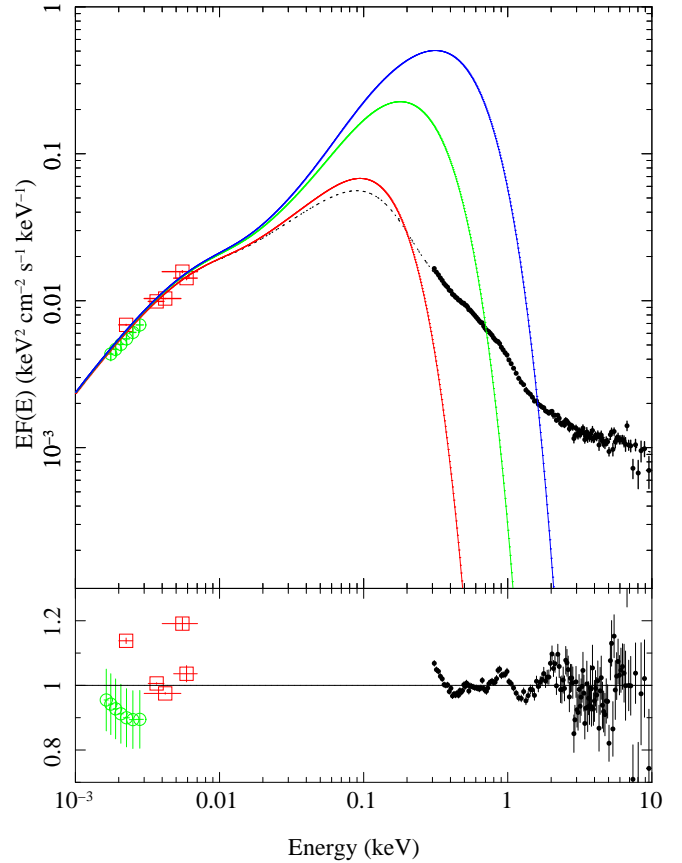}
\caption{Continuum fitting with {\sc optxagnf} to the NLS1
PG1244+026. All data and models are corrected for small amounts of
absorption due to neutral interstellar absorption and reddening by
dust. The dotted black line indicates the best-fitting {\sc optxagnf}
model for a black hole spin fixed at
zero, with mass fixed at the maximum of $2\times 10^7\Msol$. The
red/green/blue solid lines show a comparison pure disc 
spectrum (no soft excess or hard tail)
for spin of $0, 0.9$ and $0.998$, respectively.
Clearly the data strongly favour a low
spin black hole for this assumed (maximum) mass.}
\label{fig:pg1244_opt}
\end{figure}

\section{The spectral energy distribution of PG1244+026}

Our new XMM-Newton data were taken on 2011-12-25, and combine X-ray
spectral data from the EPIC cameras, together with simultaneous
optical and UV photometry from the Optical Monitor (OM). The
extraction of these data follows standard procedures, and is 
described in detail in J13. 

The OM data are corrected for emission line contamination using
archival optical (SDSS) and UV (HST) spectra, which leads to a
reduction in each OM filter band by $\sim 10$\%. The new photometric data
then match reasonably well to the archival SDSS optical continuum points,
corrected for emission lines, FeII blends and Balmer continuum
components (described in detail in J12a).

We use the {\sc optxagnf} model described above which is included in
{\sc xspec} (Arnaud 1996) version 12.7.1 and above. We correct for
absorption in the X-ray and optical/UV in both our Galaxy and the host
galaxy using the {\sc (z)wabs} and {\sc
(z)redden} models, with E(B-V) in {\sc (z)redden} fixed to
$1.7\times$ the X-ray column in units of $10^{22}$~cm$^{-2}$. We fix
the interstellar column through our Galaxy at $1.87\times
10^{20}$~cm$^{-2}$, but leave that of the host galaxy as a free
parameter. 

The {\sc optxagnf} model requires the mass of the black hole as an
input parameter.  The archival Sloan Digital Sky Survey (SDSS) optical
spectrum clearly shows the classic broad and narrow emission line
components (J12a) allowing an initial mass estimate to be derived from
standard scaling relations (Kaspi et al 2000).  These require the
width of the broad component of the H$\beta$ line which we derive
after subtracting a narrow component whose profile is matched to the
narrow [OIII] 5007\AA\ line profile. The resulting full-width half
maximum (FWHM) is 950 km/s, making this one of the most extreme
objects in the Narrow Line Seyfert 1 (NLS1) class, particularly at
this relatively high luminosity of $L_{5100}=4.52\times
10^{43}$~ergs~s$^{-1}$ (Boroson \& Green 1992). Together, the FWHM and
$L_{5100}$ values give a black hole mass estimate of $2.5\times
10^6\Msol$ (J12a). A conservative uncertainty of $\pm 0.5$~dex gives
$0.8 - 8.0 \times 10^6\Msol$.  This mass range means the bolometric
luminosity $L_{bol}=1.8\times 10^{45}$~ergs~s$^{-1}$ (estimated in the
spectral fits below) is a factor $17-1.7 \times$ higher than the
Eddington luminosity, $L_{Edd}$. Radiation pressure can then have a
marked effect on the dynamics of the broad line region. Taking this
into account in the scaling relations (Marconi et al 2008) increases
the initial mass estimate to $2.5 \times 10^7\Msol$.

However, we caution that these corrections for radiation pressure are
poorly known, so instead we derive independent constraints on the mass
from the X-ray variability properties. We split our 2-10~keV X-ray
lightcurve into two 40ks segments, binned on 250s, and find an average
the fractional excess variance of $\sigma/I=17.5 \pm 0.5$ per cent
(J13). Comparing this to the reverberation mapped sample of Ponti et
al (2012) gives a mass range of $0.2 - 2 \times 10^7\Msol$.  We take
this as the more robust estimate for the mass range of PG1244+026

\begin{figure*}
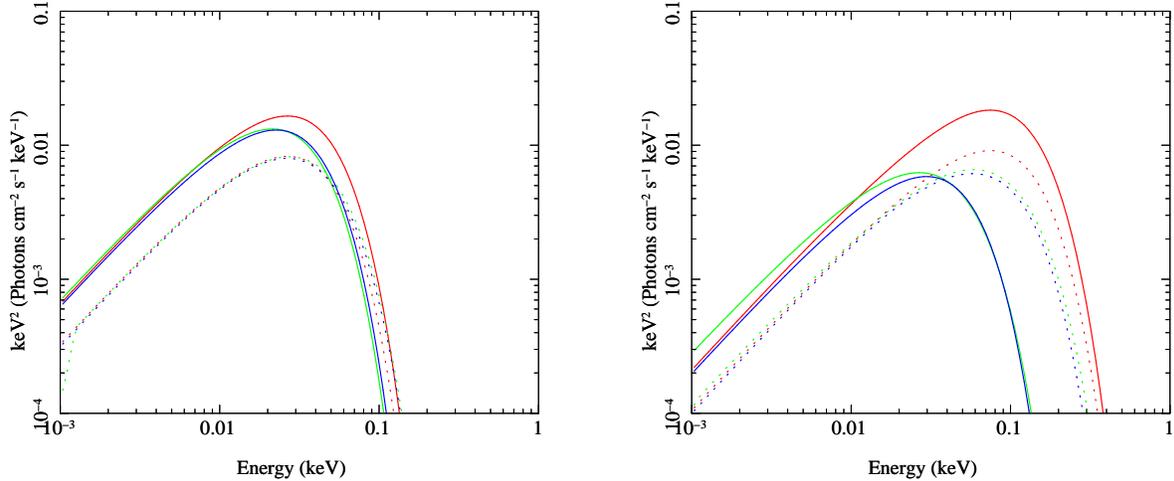

\centering
\begin{tabular}{cc}
\leavevmode  
\epsfxsize=0.45\textwidth \epsfbox{optxagn_optxconv_kerrbb_a0.ps} &
\epsfxsize=0.45\textwidth \epsfbox{optxagn_optxconv_kerrbb_a0998.ps}\\
\end{tabular}
\caption{Relativistic effects from propagation of light for spin=0
(left panel) and 0.998 (right panel) for a $10^7\Msol$ black hole
accreting at $L/L_{Edd}=0.1$. The blue line shows a pure disc spectrum
with $f_{col}=1$ using the full relativistic kernel (kerrbb) while the
green line shows our model with $f_{col}=1$ convolved with the
relativistic kernel for an inner and outer radius of $1-2 R_{isco}$
(kerrconv: green line), while the unconvolved spectrum is shown in red.
The solid line shows results for an
inclination of $0^\circ$, while the dotted line shows $60^\circ$,
assuming isotropic emission. Our approximation (green) matches very well to
the full calculations (blue) over the entire spin and inclination range
covered here.}
\label{fig:conv}
\end{figure*}

Based on the above, we fix the mass at the upper limit of $2 \times
10^7\Msol$, which gives the minimum $L/L_{Edd}$ and hence the minimum
predicted disc temperature. The fit parameters are detailed in Table
\ref{tab:pg_optxagnf}. Figure \ref{fig:pg1244_opt} (top panel) shows
the resulting {\sc optxagnf} model fit (dotted black line) to the
absorption corrected data assuming a black hole spin of $a_*=0$.  The
red line shows the model assuming all the energy thermalises in the
disc.  This is very similar to the model fit to the data apart from in
the soft and hard X-ray regimes, showing that the energy dissipated in
these components is only a small fraction of the total inferred
bolometric flux. Increasing the spin to $a_*=0.9$ (green) and $0.998$
(blue) increases the disc temperature sufficiently to strongly
overpredict the observed soft X-ray flux.  The fit residuals (bottom
panel) are dominated by the XMM-Newton OM points, specifically the V
and UVM2 wavelength filters. These discrepancies could be due to residual
aperture/emission line effects despite our efforts to correct for them
(see also J13). Removing these points does not significantly affect
the fit, so we include them as these are truely simultaneous (unlike
SDSS).

The {\sc optxagnf} model does not include relativisitic
effects, nor inclination corrections (it assumes an inclination of
$60^\circ$), nor does it account for advection. We include these
factors below.

\section{A relativistic AGN continuum model: {\sc optxconv}} 

We extend the {\sc optxagnf} model to incorporate an inclination
dependence and relativistic effects.  Inclination affects the observed
disc emission due to the angular dependence of the radiation. Full
radiative transfer photosphere models (e.g. those used by Davis \&
Laor 2011) show that the optical emission is fairly isotropic for
inclination angles $i<60^\circ$, so we normalise our model by $\cos
i/\cos 60^\circ$ to account for this. 

Inclination also affects the observed appearance of the disc through
relativistic effects. In principle we could convolve the colour
temperature corrected blackbody spectrum at each radius with the
appropriate relativistic smearing kernel at that radius, and integrate
over the whole disc. However, this is very inefficient in terms of
computational time. Instead, we use the fact that most of the disc
emission arises from radii less than twice that of the innermost
radius, so smear the entire disc spectrum with the relativistic kernel
for the appropriate spin and inclination ({\sc kerrconv}: Brenneman \&
Reynolds 2006), with
emissivity fixed at 3 and with inner and outer radii fixed at
$R_{cor}$ and $2R_{cor}$, respectively. The energy released below
$R_{cor}$ is assumed to have a constant spectrum as a function of
radius, so we convolve the soft excess and high energy tail by the
relativistic kernel with inner and outer radius fixed at $R_{ISCO}$
and $R_{cor}$, respectively. Thus the entire model can be built
using only two convolutions.

\begin{table*}
\begin{tabular}{l|l|l|l|l|l|l|l|l}
 \hline
   \small source & D & M & i & {\sc kerrbb} spin & {\sc bhspec} spin & {\sc optxconv} spin
   & {\sc optxconv} $R_{cor}$ &  ObsID\\
\hline
LMC X-3 & 52 & 7 & 67 & $0.11\pm 0.01$ & $<0.006$ & $0.12\pm 0.03$ & $<6.3$ &
   20188-02-11-00 \\
GRO J1655-40 & 3.2 & 7 & 70 & $0.601\pm 0.002$ & $0.639\pm 0.002$ &
   $0.61\pm 0.01$ & $4.7\pm 0.1$ & 10255-01-23-00\\
XTE J1550-564 & 5.3 & 10 & 72 & $0.097\pm 0.06$ & $0.115\pm 0.03$ &
   $0.11\pm 0.03$ & $7.2_{-0.5}^{+2.3}$ &  30435-01-12-00\\
\hline
\end{tabular}
\caption{Comparison of the {\sc optxconv} spin results with those from
disc continuum fitting using {\sc kerrbb} and {\sc bhspec} (taken
from multiple disc dominated spectra in Davis et al 2006). All
system parameters of distance (D), mass (M), and inclination (i) are
held fixed for all fits. The {\sc optxconv} model is fit to only one
RXTE dataset and has
fixed $f_{pl}=1$ abd $\Gamma=2.1$.  Errors are purely
statistical. In practice, uncertainties in spin estimates are
dominated by systematic uncertainties in D,M,i. }
\label{tab:bhb}
\end{table*}

Figures \ref{fig:conv}a,b show that this is a good approximation by comparing
the results of our code, modified using the approximate smearing
functions (green line) with disc models which incorporate the full
relativistic kernel (blue line: {\sc kerrbb}: Li et al 2005). We fix
both models to $f_{col} = 1$ and use isotropic emission so they are
directly comparable. The red line shows the extent of the relativistic
effects by comparing to our model without the relativistic
smearing. We show a sequence of models for inclination of $0^\circ$
(solid lines) and $60^\circ$ (dotted lines) for spin $a_*= 0$ (left
panel) and $0.998$ (right panel). It is clear that the relativisitic
effects make most difference for high spin and low inclination (red
line most different from green and blue), and that the pure disc
results are recovered by our model for the entire range of inclination
and spin considered here i.e. good match of green and blue lines in
all cases.

\subsection{Fitting to black hole binaries}

The {\sc optxconv} model then will reproduce all the well known
results from {\sc kerrbb} fits to black hole binaries (BHB) as Fig
\ref{fig:conv} shows it reproduces the shape of {\sc kerrbb}, and its
analytic expression for $f_{col}$ is that of Davis et al (2006)
i.e. gives $~1.7-1.9$ for typical BHB mass and mass accretion rates,
as used in all {\sc kerrbb} fits.  However, we show this explicitally
by picking three black holes with solid spin estimates from disc
continuum fitting, namely GRO J1655-40, XTE J1550-564 and LMC
X-3. These were all considered in Davis et al (2006), but there have
also been multiple other studies of spin from their disk dominated
spectra (e.g. GRO J1655-40: Shafee et al 2006; XTE J1550-546: Steiner
et al 2011; LMC X-3: Steiner et al 2010). These previous spin results
are shown in Table \ref{tab:bhb}. We pick specific RXTE ObsID's from
the disc dominated states of Davis et al (2006) which lie on the
constant radius part of the luminosity-temperature plot. The spin
results from {\sc optxconv} are given in Table \ref{tab:bhb} and are
all consistent with those from Davis et al (2006) for the same assumed
parameters (distance, black hole mass and inclination). We note that
these are not always consistent with the iron line spins, nor with the
spins derived from disc continuum fitting in data which are not
dominated by the disc (as required for an iron line profile)
e.g. Miller et al (2009) derive $a_*=0.92\pm 0.02$ from both disc and
iron line in a non-disc dominated spectrum of GRO J1655-40. Non-disc
dominated spectra do not show a robust luminosity-temperature relation
so cannot be used for reliable disc continuum fitting (Kubota \& Done
2004). Using the iron line alone gives a black hole spin which is
higher than the disc continuum fits (see introduction in Kolehmainen
\& Done 2010).  The line can easily give lower spin than the disc
dominated continuum fits if the disc does not extend down to the ISCO
in the states where the line is prominant. However, significantly
higher spin from the iron line means that one (or both) of the disc
(more likely the system parameters of mass, inclination and distance)
or iron line models is wrong.

Interestingly, both GRO J1655-40 and XTEJ1550-564 show
significant pairs of high frequency QPO's at 300/450~Hz, and
180/280~Hz respectively (Belloni, Sanna, \& Mendez 2012). Assuming
that the highest frequency of the pair represents the Keplarian
frequency at the last stable orbit gives a lower limit on spin of
$a_*>0.4$ (GRO J1655-40) and $>0.3$ (XTE~J1550-564) using the mass
estimates of Table \ref{tab:bhb}, but these are not very
constraining. 

\subsection{Fitting {\sc optxconv} to PG1244+026}

\begin{figure}
\epsfxsize=0.5\textwidth \epsfbox{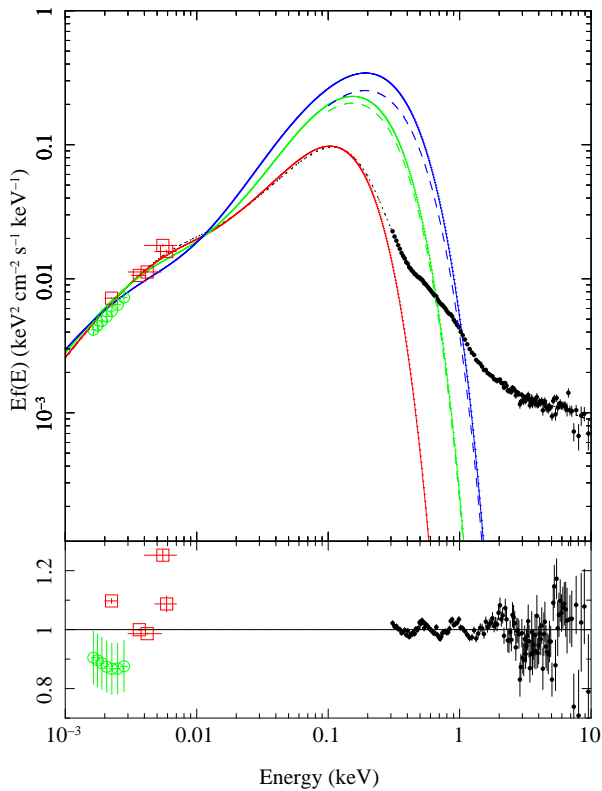}\\
\caption{Continuum fitting with {\sc optxconv} to the NLS1 PG1244+026.
All data are corrected for small amounts of absorption due to neutral
interstellar absorption and reddening by dust. The dotted black line
indicates the best-fitting model (see Table \ref{tab:pg1244} for the
parameters) for a black hole spin fixed at zero, and inclination of
$30^\circ$, which gives a best fit black hole mass of $1.1\times
10^7\Msol$ and $L/L_{Edd} = 0.8$. The red/green/blue solid line shows
a comparison of the pure disc emission (no soft excess or hard tail)
for $a_*=0, 0.9$ and $0.998$.  The dashed green and blue lines show
the reduction in inner disc emission expected from advection of
radiation for the superEddington flows at $a_*0.9$ and $0.998$,
respectively.  Clearly the data rule out the presence of a maximally
spinning black hole for this mass and inclination even including the
effects of advection.}
\label{fig:pg1244_conv}
\end{figure}

The red solid line in Figure \ref{fig:pg1244_conv} shows the new model
fit to the data from PG1244+026. We assume an inclination of $\sim
30^\circ$ as an obscuring torus probably removes all type I AGN from
$60-90^\circ$, and fix the black hole spin at $a_*=0$. This gives a
best fit black hole mass of $\sim 10^7\Msol$, and $L/L_{Edd}\sim 0.85$ (see
Table 3). Similarly to Figure \ref{fig:pg1244_opt}, we show the pure
disc spectrum as a dotted red line, and a sequence of pure disc models
with $a_*=0.9$ (green) and $0.998$ (blue). Again, the high spin
solutions strongly overpredict the observed soft X-ray flux.

However, the enhanced soft X-ray emission that derives from the disc
extending down further into the gravitational potential means that two
high spin models give $L/L_{Edd} = 2.2$ and $4.7$,
respectively. Advection can become important at these high
luminosities, with some fraction of the radiation being carried along
with the flow rather than being radiated (Abramowicz et al 1988),
suppressing the disc emission from the innermost radii. The best model
calculations of this effect (including the vertical as well as radial
structure of the disc) show the total disc emission is not affected
below $L_{Edd}$ (Straub et al 2013), but reduces to $L/L_{Edd}\sim
2.0$ and $\sim 3.5$ for $\dot{M}/\dot{M}_{Edd} = 2.2$ and $4.7$
(Sadowski 2011, their Figure 4.11). The green and blue dashed lines in
Figure \ref{fig:pg1244_conv} show the effect of this
on the predicted inner disc emission. Plainly this is insufficient
to reduce the soft X-ray flux for the high spin models to a level
compatible with the observations, ruling out a high spin scenario for
this mass and inclination.

\begin{figure}
\epsfxsize=0.5\textwidth \epsfbox{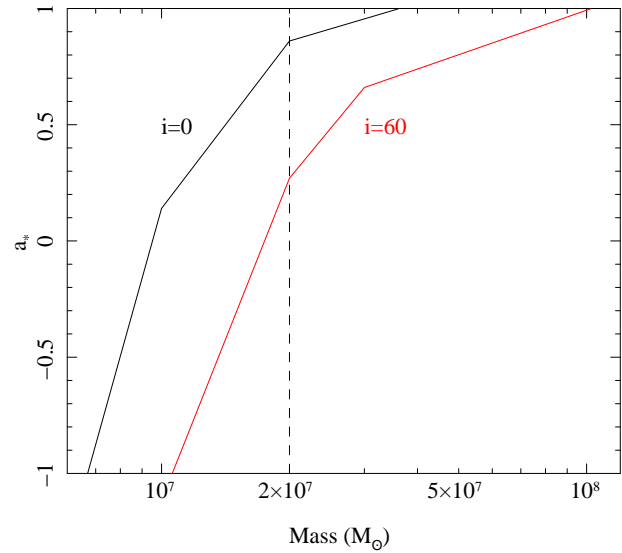}\\
\caption{The best fit mass and spin values for $i=60^\circ$ (red) and
  $i=0^\circ$ (black). The maximum mass of $2\times 10^7 M_\odot$ is
  indicated by the dashed vertical line.}
\label{fig:spin_mass}
\end{figure}

The correlated parameters are mass, spin, inclination and
$L/L_{Edd}$. We explore this parameter space by fixing inclination at
$0^\circ$ and $60^\circ$, and trace out the best fit mass for all
spins from $-1$ to $0.998$ (see Fig \ref{fig:spin_mass}). The
corresponding $L/L_{Edd}$ increases systematically with decreasing
mass and spin, with maximum of $0.85$ for $0^\circ$ and $1.3$ for
$60^\circ$, so none none of the best fit solutions are strongly
superEddington. However, at maximal spin, $L/L_{Edd}$ is much lower,
at $\sim 0.1$ and $0.3$ respectively for $i=0^\circ$ (mass $3.6\times
10^7 M_\odot$) and $i=60^\circ$ (mass of $10^8$), respectively. This
is within the range of normal Quasar $L/L_{Edd}$ (Boroson 2002), yet
the spectral energy distribution is very different to that of the
standard quasar template spectrum (e.g. Elvis et al 1994). In our
opinion this makes these higher mass/spin solutions less plausible. We
note that our maximum mass of $2\times 10^7M_\odot$ requires spin
$<0.86$ even at $i=0^\circ$.

\begin{table*}
\begin{tabular}{lc|l|l|l|l|l|l|l|l}
 \hline
   \small $N_H$ ($10^{20}$~cm$^{-2}$) & $M\ (10^7\Msol)$ &
$\log L/L_{Edd}$ & $r_{corona}$ ($R_g$)&
$kT_e$ (keV)& $\tau$ & $\Gamma$ & $f_{pl}$ & $\chi^2/\nu$ \\ 
\hline
$3.7\pm 0.2$ & $1.08\pm 0.01$ & $-0.08\pm 0.02$ & $13.0\pm 0.3$ &
$0.19\pm 0.01$ & $20.0_{-0.7}^{+1.3}$ & $2.34\pm 0.02$ &
$0.35\pm 0.02$ & 1640/1067\\
\hline
\end{tabular}
\caption{Details of the {\sc optxconv} fit to PG1244+026 shown in Fig
  \ref{fig:pg1244_conv}. The remaining paramaters of inclination and
  black hole spin are fixed at $30^\circ$ and $a_*=0$, respectively.}
\label{tab:pg1244}
\end{table*}

\section{Comparison with reflection spin estimators for PG1244+026}

There is an iron line in the X-ray data, but there is not sufficient
signal-to-noise in the 3-10~keV bandpass to use this to tightly
constrain the black hole spin (see J13 for details). Hence we cannot
currently compare the two techniques in this object.

However, reflection may also form the soft X-ray excess. This is an
alternative model to the one used here where we assumed that the
soft excess was a true additional component. 
Fitting the entire X-ray spectrum of PG1244+026 with ionised 
reflection models strongly requires high spin (Crummy et al 2006; J13), 
in conflict with the disc continuum fits above. However, the lack of 
correlated variability of the soft X-ray excess with that of the
hard power law argues against a reflection origin for the majority of the
soft X-ray excess in this object (J13), so the black hole spin estimate
from reflection fits to the soft X-ray excess are probably not 
valid in this object. 

\section{Discussion}

This is the first demonstration that black hole spin in a NLS1 can be
constrained from disc continuum fitting. The only previous application
of this technique in AGN was for a very high mass black hole accreting
at moderate rate ($L/L_{Edd}\sim 0.1$) where the disc peak is resolved
in the observable optical/UV due to the high source redshift of $z =
1.66$ (Czerny et al 2011). Instead, our study uses the lack of an
observed peak in the soft X-ray range to constrain spin in
a local, much lower mass black hole, higher mass accretion rate AGN. These NLS1
are an important class of AGN as these are typically the systems where
extreme relativistic effects requiring high spin and gravitational
lightbending are claimed from the shape of the reflected iron line
emission during low flux episodes, where the spectra appear reflection
dominated. These are interpreted in the lightbending model as due to
the continuum source dropping in height so that strong lightbending
both suppresses the intrinsic power law emitted towards the observer,
and enhances the illumination of the very innermost parts of the disc,
leading to a strongly centrally peaked emissivity (Fabian et al 2004;
Miniutti \& Fabian 2004). The higher flux spectra from these objects
are much less reflection dominated and their reflection emissivity is
less centrally concentrated, consistent with a larger source height
where lightbending is less effective.

Our low spin result for PG1244+026 is in sharp contrast with the high
(almost maximal) spin required to fit the low flux episodes of the
NLS1 1H0707-495 (Fabian et al 2004; 2009). While these are different
objects, they have similar mass and mass accretion rates, and the
X-ray spectrum of 1H0707-495 in its high flux episodes is remarkably
similar to that of PG1244+026. Cosmological simulations of the
co-evolution of black holes in AGN and their host galaxies predict
that low mass black holes should all have similar
spins as they are built from the same process (gas accretion rather
than merging black holes). These should all have high spin if the
accretion angular momentum direction is prolonged, or all low if the
accretion is chaotic (e.g. Fanidakis et al 2011). Hence our result
implies that either our disc continuum spin estimate is wrong, or the
reflection dominated interpretation of the low flux state spectra are
wrong, or that our understanding of the cosmological evolution of
black hole spin is wrong. We discuss each of these possibilities in turn below.

The disc continuum fit could underestimate spin if the mass were
significantly underestimated. This does not seem likely given the very
narrow line widths in this object and its rapid X-ray variability (see
Section 3). Alternatively, advection could become much more important
if the black hole mass were instead towards the lower limit of the
probable range.  While a very super-Eddington mass accretion rate would
not be an issue for a single object, we note that the high $L/L_{Edd}$
sample of J12a,b all have similar X-ray spectra and masses, so
probably will all have similarly low spin constraints. With the
current models these AGN all have $L/L_{Edd}\sim 1$, following
smoothly on from the other two subsamples which have $<L/L_{Edd}>\sim
0.2$ and $\sim 0.05$ (J12a,b, D12). If instead all these objects were
super-Eddington there would be a deficit of systems with mass accretion
rate around $L/L_{Edd}\sim 1$. The only other possibility is that the
disc models themselves are wrong, despite being solidly tested in the
BHB systems. The most significant difference made by increasing the mass
is that the disc temperature decreases. This means that the disc can
power a UV line driven wind (Proga et al 2000; Risaliti \& Elvis 2010)
and mass loss in this wind could be substantial enough to change the disc
structure (Laor \& Davis 2013).

Alternatively, the high spin derived from X-ray reflection models for
the X-ray low states of NLS1 such as 1H0707-495 could be
overestimated. This can be the case if most of the soft X-ray excess
is a true additional continuum but is erroneously fit with reflection
models. The strong relativistic effects required to smear the
predicted soft X-ray line emission into the observed continuum then
drive the fit, as the statistics at low energies are much better than
at the iron line e.g. most of the objects in Crummy et al (2006)
require high spin at high significance. However, the low flux spectra
which most strongly require high spin are often fit with two
reflectors, one for the soft X-ray excess, and another for the iron
line but the iron line alone strongly requires high spin in 1H0707-495
(Fabian et al 2009). Instead, the curvature around the iron line
could be due to absorption rather than to relativisitic effects
(Turner et al 2007; Miller et al 2007). This appears to require a
fine-tuning of the geometry (Zoghbi et al 2011) but these
extreme line profiles are only seen in a subset of NLS1, those where
the flux drops dramatically (Gallo 2006) which may select
objects where the line of sight is directly down the wind.
Alternatively, the reflection spectrum could itself be distorted if
the disc photosphere is strongly turbulent and/or dominated by Compton
scattering, as predicted for the inner
regions of the disc in the failed wind/hitchhiking gas model of
Risaliti \& Elvis (2010).

Finally, the cosmological spin evolution could be wrong. This seems
almost certain as the models only include spin up/down from accretion
and black hole mergers, yet these objects also can power jets at some
stages of their active lifetimes. The jet may remove angular momentum
from the black hole if it is powered by the Blandford-Znajek
process though this is also still controversial as the jet could also
be powered simply by accretion e.g. Russell, Gallo \& Fender (2013).

\section{Conclusions}

This paper demonstrates that disc continuum fitting can constrain
black hole spin in AGN, though there are still systematic
uncertainties from black hole mass determinations which require
reverberation mapping to substantially reduce. Applying the
continuum fitting technique to a sample of disc dominated AGN to
derive the spin distribution will give new insight into current
controversies which beset the interpretation of X-ray spectra. More
fundamentally, measuring black hole spin reliably would give a test of
the origin of jet power, and shed new light on the nature of the
accretion powered growth of super-massive black holes across cosmic
time.

\section*{Acknowledgements}

CD acknowledges a conversation with Ric Davies at MPE where in
describing the need for new data to use the disc continuum fitting
technique in AGN, it became clear that this was already feasible.


\label{lastpage}
\end{document}